\documentclass[10pt]{iopart}
\usepackage{epsfig,graphicx}

\begin{document}

\title{Adiabatic spin cooling using high-spin Fermi gases}

\author{M. Colom\'e-Tatch\'e$^1$, C. Klempt$^2$, L. Santos$^1$, and T. Vekua$^1$}
\address{$^1$ Institut f{\"u}r Theoretische Physik, Leibniz Universit\"at Hannover, Appelstr. 2 D-30167,
Hannover, Germany}

\address{$^2$ Institut f\"ur Quantenoptik, Leibniz Universit\"at Hannover, 30167~Hannover, Germany}

\newcommand{\beqa}{\begin{eqnarray}}
\newcommand{\eeqa}{\end{eqnarray}}

\newcommand{\om}{\omega}

\begin{abstract}
Spatial entropy redistribution plays a key role in the adiabatic cooling of ultra-cold lattice gases. 
We show that high-spin fermions with a spatially variable quadratic Zeeman coupling may 
allow for the creation of an inner spin-$1/2$ core surrounded by high-spin wings. The latter are always more entropic 
than the core at high temperatures, and remarkably at all temperatures in the presence of frustration.
Combining thermodynamic Bethe Ansatz with local density approximation, we study the 
spatial entropy distribution for the particular case of one-dimensional spin-$3/2$ lattice fermions in the Mott-phase. 
Interestingly, this spatially-dependent entropy opens a possible path for an adiabatic cooling technique 
which contrary to previous proposals, would specifically target the spin degree of freedom. We 
discuss a possible realization of this adiabatic cooling, which may allow for a high-efficient entropy decrease 
in the spin-$1/2$ core and help accessing antiferromagnetic order in experiments on ultracold spinor fermions.

\end{abstract}

\maketitle

\section{Introduction}

Ultra-cold atoms in optical lattices offer an extraordinary controllable scenario for the study of 
strongly-correlated systems~\cite{Lewenstein2006,Bloch2008}, exemplified by 
the observation of the superfluid to Mott-insulator (MI) transition in 
ultra-cold bosons~\cite{Greiner2002}. 
Remarkable progress has been achieved in lattice fermions as well, allowing for 
the precise analysis of the Fermi-Hubbard model, a key model in condensed-matter physics  
of particular relevance in the study of high-temperature superconductivity~\cite{Hofstetter2002}. 
Exciting recent experiments have reported the realization of the metal to MI transition 
in two-component fermions~\cite{Jordens2008,Schneider2008}. 
Due to super-exchange, the MI phase of spin-$1/2$ fermions 
is expected to exhibit a magnetic N\'eel (antiferromagnetic) ordering.
On-going experiments are already very close to reach the mean-field entropy per particle, 
$s$, for N\'eel ordering in a 3D cubic lattice ($s/k_B=\ln2$)~\cite{Jordens2010}. 
However quantum corrections reduce the critical $s$  down to $s_N/k_B\simeq 0.35$~\cite{Werner2005,Koetsier2008}. 

Reaching such an extraordinary low entropy constitutes nowadays a major challenge, which demands 
novel types of cooling especially designed for many-body systems in optical lattices\cite{McKay}. A number of 
cooling proposals have been recently suggested~\cite{Popp2006,Werner2005,Dare2007,Bernier2009,Ho2009a,Ho2009b,Catani2009,Heidrich-Meisner2009}, 
most of them based on the redistribution of entropy within the trap, where 
certain regions act as entropy absorbers from the region of interest, i.e. a Mott insulator at the trap center.

Interestingly, spin degrees of freedom may be employed for designing cooling techniques 
resembling adiabatic demagnetization cooling in solid-state physics~\cite{Tishin2003}. 
In this method, a decrease in the strength of an externally applied magnetic field allows 
the magnetic domains of a given material to become disoriented. If the material is isolated, 
temperature drops as the disordered domains absorb thermal energy in order to perform their reorientation. 
In cold atoms, this technique was pioneered in Chromium experiments, 
where the spin-flip mechanism was provided by dipole-dipole interactions~\cite{Fattori2006}. 
Recently, a novel demagnetization cooling mechanism has been proposed for two-component fermions 
based on time-varying  magnetic field gradients~\cite{Medley2010}. In that method, scalar domains are 
cooled by transferring particle-hole entropy into magnetic entropy in overlapping regions between the two components. 
Note, however, that gradient cooling does not address cooling of the spin degrees of freedom, contrary to the method discussed below.

In this paper, we study the spatial entropy distribution of multi-component spin-$S$ fermions with 
an inhomogeneous quadratic Zeeman effect~(QZE), and how this spatially-dependent entropy profile may 
be employed for designing an adiabatic cooling method which specifically targets the reduction of spin entropy.
We show in particular, that an inhomogeneous QZE may lead to an effective pseudo-spin-$1/2$ core surrounded 
by spin-$S$ fermions at the wings. We show that, remarkably, the spin-$S$ wings act as entropy absorbers all the way 
to vanishing temperatures in the presence of frustration. We illustrate the idea with the specific example of one-dimesnional
(1D) spin-$3/2$ lattice fermions in the Mott phase. Interestingly, the adiabatic growth of the spatially dependent QZE 
combined with spin redistribution via spin-changing collisions may open promising perspectives towards adiabatic cooling 
of the spin-$1/2$ core, which may in this way enter the antiferromagnetic spin coherent regime.

\section{Entropy and frustration}

In the following we consider multi-component fermions loaded in an optical lattice.
Increasing the number of spin components from $2$ to $N$ (effective spin $S=(N-1)/2$) increases 
the capacity to allocate larger entropy per spin at high $T$ by a factor of $\ln N/\ln 2$. 
This guarantees that at high $T$, well over the N\'eel ordering, spatial regions with a
larger effective $S$ act as entropy absorbers in the cooling process shown below. 
However, this simple argument does not apply at low $T$ if the system acquires conventional N\'eel ordering~\cite{footnote1}.
The entropy of the Heisenberg antiferromagnet (HAF) in $d$ dimensions
scales for $T\ll T_N$ (N\'eel temperature~\cite{footnote1}) as $s\sim S^{-d}$, and hence the entropy of a spin-$S$ system decreases as compared to that of a spin-$1/2$ HAF.
This is clear since larger spins attach more to the N\'eel direction, and hence the density of states is smaller at low $T$ leading to a lower entropy. 

The situation is reversed in frustrated systems, which present a large degeneracy of classical ground states with many branches of soft excitations at low $T$. 
One arrives at the simplest frustrated large-$S$ model starting from the $SU(N)$ symmetric Hubbard model
\begin{equation}
H=-t\sum_{m,<i,j>} (c^{\dagger}_{m, i}c_{ m, j}+h.c.)+{U}/{2} \sum_{i} n_{i}^2,
\label{eq:H}
\end{equation}
where $m=(-S,\dots,S)$, 
$c_{ m, i}$ annihilates fermions with spin $m$ in the site $i$, 
$n_i=\sum_m n_{m,i}=\sum_m c^{\dagger}_{m, i}c_{ m, i}$, and $t$ and $U$ are the hopping and interaction 
coupling constants, respectively. In the strong-coupling limit, $U\gg |t|$, and retaining one fermion per site, one can derive the effective permutation model in second order of perturbation theory,
\begin{equation}
\label{permutation}
H_0={J}/{2}\sum_{<i,j>}P_{i,j},
\end{equation}
with $P_{i,j}$ being the permutation operator and $J=4t^2/U$. This model is the $SU(N)$ generalization of the HAF. It is exactly solvable in 1D~\cite{Sutherland1975} and has $N-1$ gapless spin modes, each dispersing at low momenta with velocity $2v_{1/2}/N$, where $v_{1/2}=\pi J/2$ is the spin wave velocity of the spin 1/2 HAF.
At low $T$, the entropy $SU(N)$ spin model is larger than that of spin-$1/2$ HAF, due to increasing fluctuations by 'orbital' degrees of freedom, following $s_S(T)=N(N-1)\pi T/6v_{1/2}$ \cite{Sutherland1975}.
For equivalent 2D and 3D lattice models an increase of entropy with unbinding number of degrees of freedom is also expected at low $T$. There, the classical ground state shows extensive degeneracy, and it is believed \cite{Moessner1998} that the N\'eel order does not get stabilized for $SU(N>2)$ due to high frustration~\cite{footnote2}.

Hence, due to frustration, with increasing $S$, N\'eel order gets suppressed leading to a larger entropy storage capacity at low $T$. Thus, if the Mott edges get frustrated while preserving $S=1/2$ character in the central region, {\em one can use the frustrated edges as entropy absorbers all the way from high to extremely low $T$}.

\section{1D Spin-$3/2$ fermions in the Mott-phase}

In the following, we illustrate the possibilities provided by high-spin lattice fermions 
with the specific case of a balanced mixture of spin-$3/2$ fermions in a 1D optical lattice, in which 
the number of particles $N_m$ with spin $m$ satisfies $N_m=N_{-m}$. Interparticle interactions are characterized 
by the $s$-wave scattering lengths for channels with total spin $0$ and $2$, $a_{0,2}$. 
For  $s$-wave interacting fermions $S=3/2$ is the lowest spin allowing for spin-changing collisions (which preserve the total magnetization 
but transfers atoms from $\pm 1/2$ into $\pm 3/2$ and vice versa). 
Due to the conserved magnetization the linear 
Zeeman effect does not play any role.
However, the QZE, characterized by the externally controllable constant $q$, induces a finite chirality $\tau=\frac{1}{L}[(N_{3/2} + N_{-3/2})-(N_{1/2} + N_{-1/2})]$. 

For large-enough interactions and at quarter filling (one fermion per site) the 1D system enters into the Mott insulator regime, for which the ground state properties under QZE were studied in Ref.~\cite{Rodriguez2010}. For large QZE the ground state is a pseudo spin-$1/2$ isotropic HAF. Reducing $q<q_{cr}$ ($q_{cr}=J\ln 2/2$ at the $a_0=a_2$) the system enters either a spin liquid phase (for $a_2<a_0$) or a dimerized phase (for $a_2>a_0$). 
For $a_0\simeq a_2$ (the typical situation unless $a_{0,2}$ 
are externally modified), the gap of the dimerized phase is exceedingly small, and 
hence the system behaves in practice as a spin liquid down to extremely low temperatures.
For $a_0=a_2$, in the presence of a spatially variable QZE, the model Hamiltonian becomes $H=H_0+\sum_i\mu_{m,i} n_{m,i}$, with $H_0$ given by Eq.~(\ref{permutation}) with $N=4$. For homogeneous $\mu_{m,i}$, this model is exactly solvable and its thermodynamic properties may be calculated by means of thermodynamic Bethe Ansatz. We follow the method of Ref.~\cite{Damerau2006}, based on the self-consistent solution of $14$ coupled integral equations, to obtain the corresponding free energy $f$.  
The chemical potentials for each component are $\mu_{m,i} = \mu + m^2 q_i$, where $\mu$ is the global chemical potential, and $q_i$ denotes the QZE constant at site $i$. The entropy is then given by $s=-\partial f/\partial T$ and the chirality by $\tau=-\partial f/\partial q$. 

\begin{figure}[t]
\vspace*{0.2cm}\centering\includegraphics*[width=0.9\columnwidth]{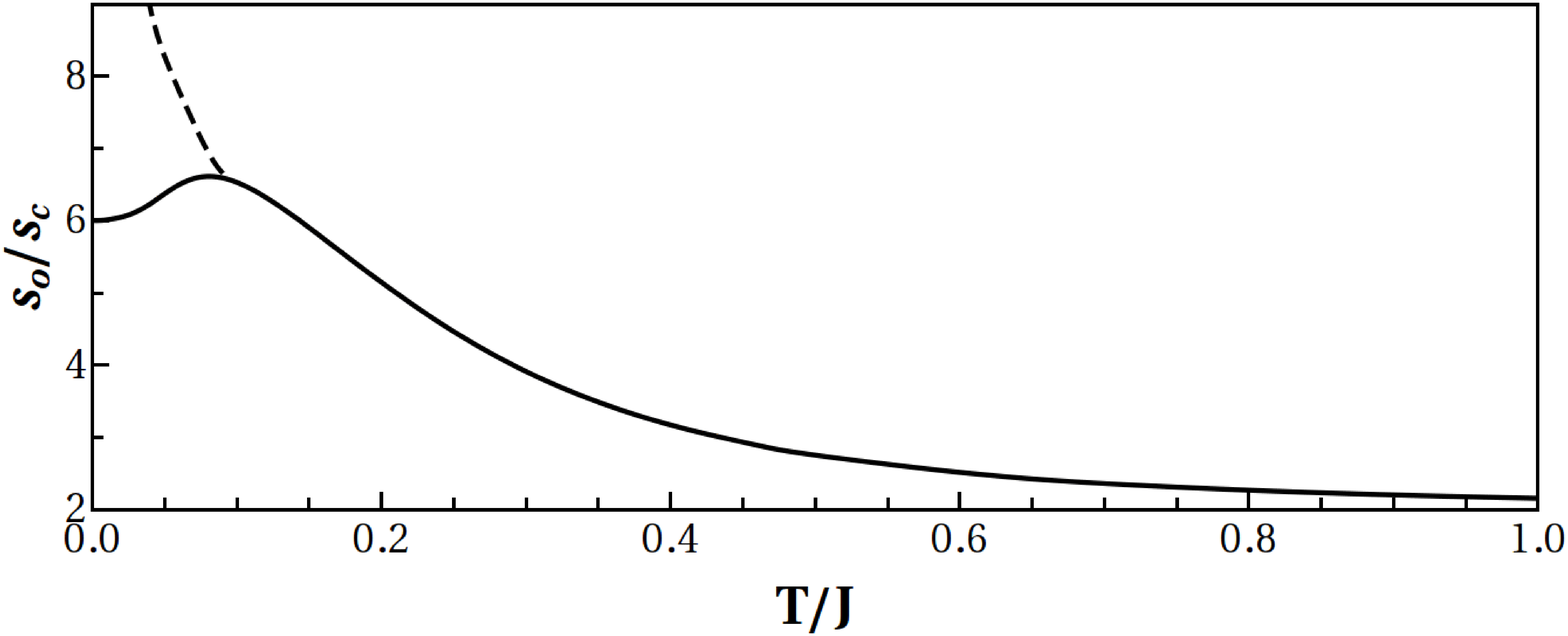}
\vspace*{-0.2cm}\caption{Ratio between the entropy per particle $s_0$ at $q=0$ (continuous line) or at $q={\tilde q}(T)$ (dashed line) (${\tilde q}(T)$ is the QZE at which entropy is largest at a given $T$), and the entropy $s_c$ at high QZE ($q=q_0$), versus the temperature $T$ in the lattice. Observe the crucial role of frustration at low $T$. If we were dealing with an unfrustrated $S=3/2$ HAF, the curve would have arrived at low $T$ at $\simeq 0.4$~\cite{Hallberg96} instead of $6$.}
\label{fig:1}
\vspace*{-0.3cm}
\end{figure}

As mentioned above, for $q>q_{cr}$ the system becomes pseudo-spin-$1/2$. Hence, the ratio between the entropy per spin 
for $q=0$~($s_0$) and that for $q>q_{cr}$ should follow the same dependence as $s_{3/2}(T)/s_{1/2}(T)$. 
We illustrate this point in Fig.~\ref{fig:1}, where
we compare $s$ at large $q=q_0=5J$ ($s_c$) to $s_0$.
Note that at large $T$ $s_0/s_c=2$, whereas  at low $T$ frustration leads to $s_0/s_c=6$.

\section{Spatially-dependent QZE}

We consider at this point the entropy and chirality profiles for the case of a non-homogeneous QZE, which may be achieved  
by means of microwave or optical techniques~\cite{Tannoudji1972,Gerbier2006,Santos2007}.
We perform local QZE approximation (similar to the local density approximation standard in trapped gases), i.e. we solve for 
the free energy at different positions ${\mathbf x}$ by varying $q({\bf x} )$. The local QZE approximation demands 
a sufficiently slow variation of the QZE at the scale of the inter-site spacing. In this way we can evaluate the 
entropy profile inside the Mott insulator region. Note finally that although the calculation is done for the case $a_0=a_2$, 
the conclusions may be extended to the actual case in which $a_{0}$ and $a_{2}$ are slightly different, 
and spin redistribution via spin-changing collisions occurs.

\begin{figure}[t]
\centering
\includegraphics*[width=0.49\columnwidth]{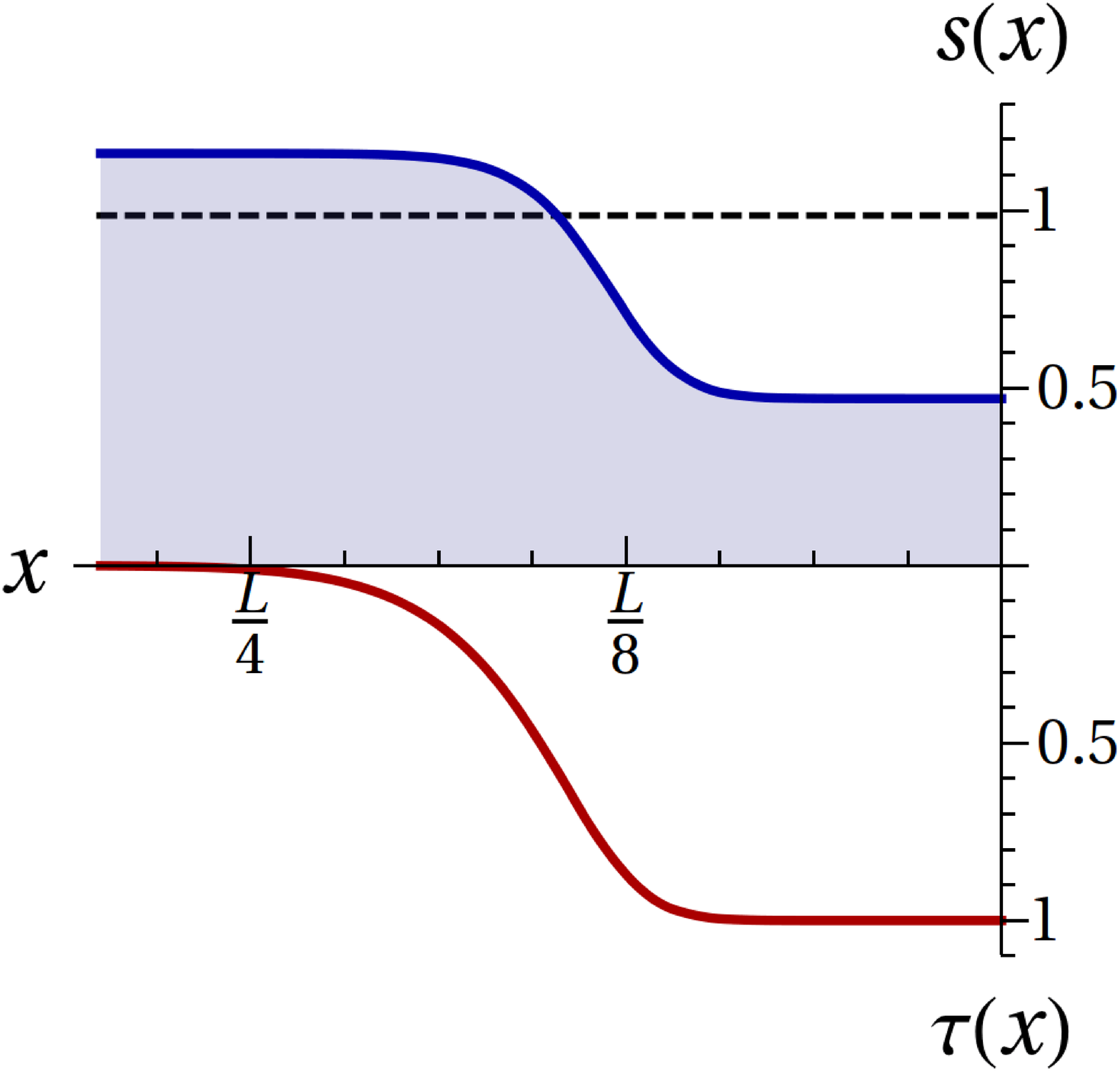}
\includegraphics*[width=0.49\columnwidth]{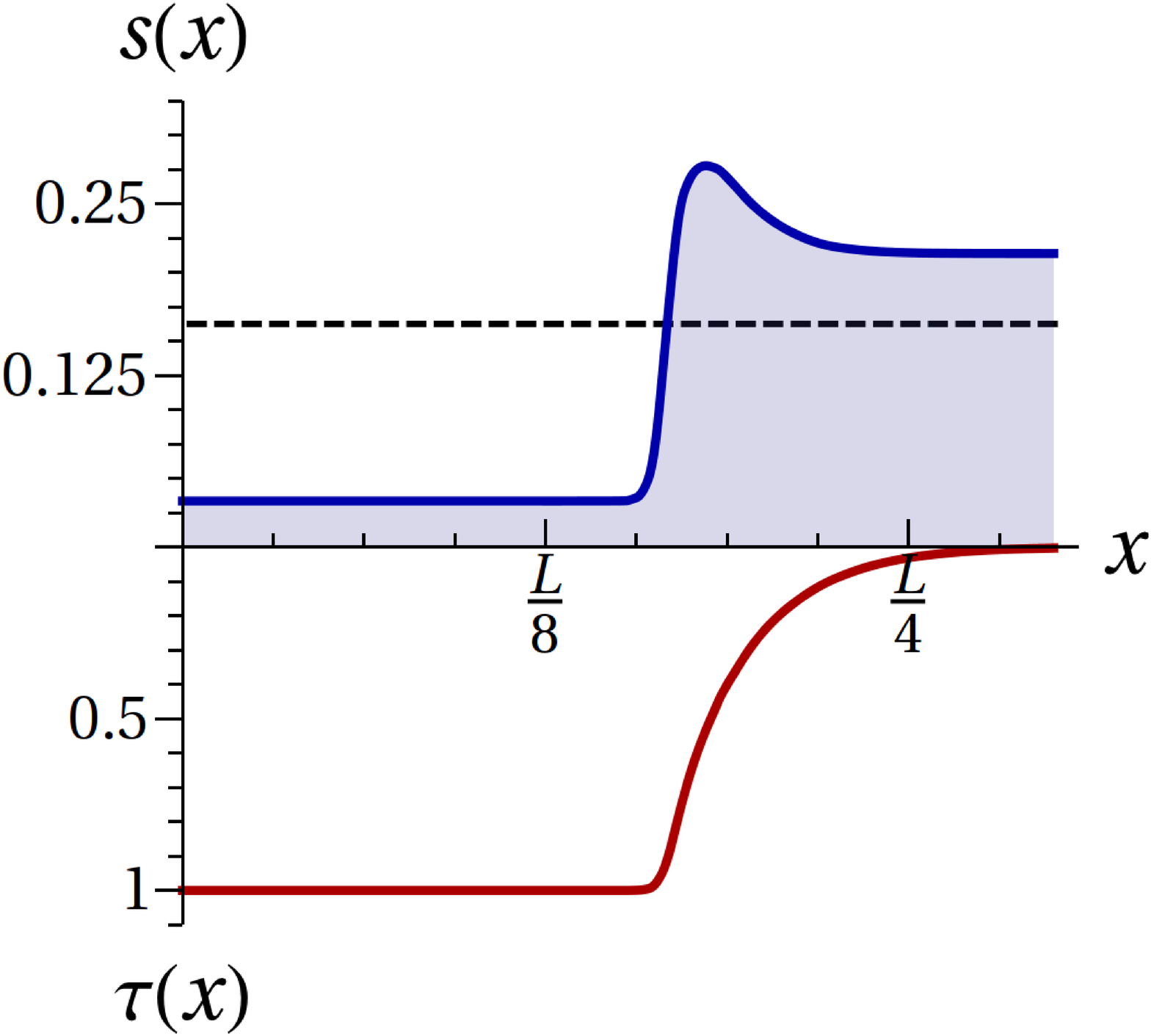}
\vspace*{-0.2cm}
\caption{Entropy per particle (in $k_B$ units) and chirality profiles for $L=120$ particles for a Gaussian QZE profile $q(x)=q_0\exp\left({-{100 x^2}/{L^2}}\right)$, with $q_0=5J$.
Dashed lines indicate the initial entropy prior to the switching of the inhomogeneous QZE profile. Before the lattice loading, $T_i/T_F=0.1$ (left) and $0.016$ (right). Note that for $q>q_{cr}$, $\tau\simeq 1$ and the system retains a spin-$1/2$ character. Note also how the entropy gain in the Mott core is larger for small $T$. The entropy bump for small $T$ at $q=q_{cr}$ reflects the Van-Hove singularity at the bottom of the depleted Hubbard band.}
\vspace*{-0.4cm}
\label{fig:2}
\end{figure}

Figure~\ref{fig:2} shows chirality and entropy profiles. Note that the entropy per site (i.e. per particle) 
is significantly larger at the Mott wings than at the center.
Hence, if the total entropy is conserved in a process in which an initially 
spatially-independent entropy (deshed line on Fig.~\ref{fig:2}) is brought to the profile of Fig.~\ref{fig:2},   
the outer regions will remove entropy from the central core.
This process is even more effective at low $T$.
One may indeed estimate the entropy reduction (for $T\to 0$) at the Mott center
for model~(\ref{permutation}) in 1D, when considering an initial uniform spin-$1/2$ system and a
final step-like distribution with a spin-$1/2$ core and spin $S=(N-1)/2$ wings:
\begin{equation}
\label{gain}
\gamma\equiv s_i/s_c= [(1-L_0/L)N(N-1)/2+L_0/L],
\end{equation}
where $L_0/L$ is the ratio of the number of sites in the spin-$1/2$ core to the total number of sites,and $s_i$  ($s_c$) is the initial (final) entropy per particle at the center.
The gain is hence much bigger than the one at $T\gg T_N$, which is obtained after one substitutes $N(N-1)/2$ by $\ln N/\ln 2$ in Eq.~(\ref{gain}).


\section{Adiabatic cooling}

\begin{figure}
\centering
\includegraphics*[width=0.8\columnwidth]{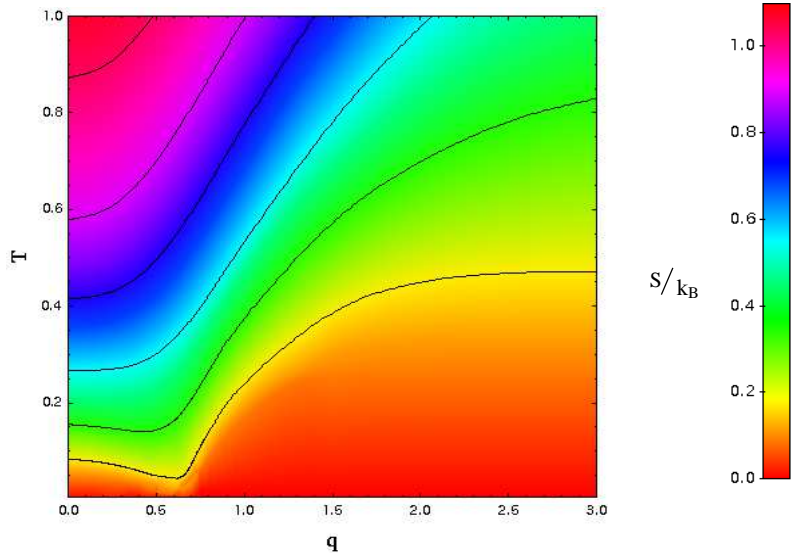}
\vspace*{-1.5cm}
\caption{Sliding down the isentropic curves (solid curves) with reducing uniformly QZE induces the reduction of the temperature in the lattice at the second stage of the proposed 
adiabatic cooling scheme.
}
\vspace*{-0.3cm}
\label{fig:Isentropes}\end{figure}

\begin{figure}
\centering
\includegraphics*[width=0.9\columnwidth]{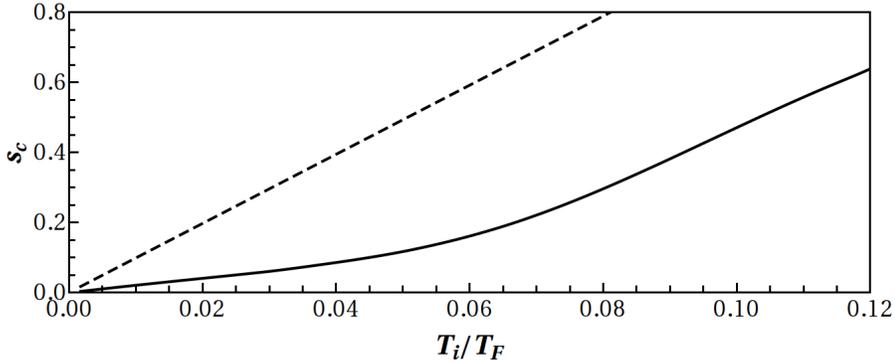}
\vspace*{-0.2cm}
\caption{Entropy at the center $s_c$ (in $k_B$ units) as a function of the temperature before lattice loading, $T_i/T_F$, for the inhomogeneous $q(x)$ profile of Fig.~\ref{fig:1}~(solid) and for a homogeneous one $q=q_0$~(dashed), for which the whole system retains a spin-1/2 character.
In order to associate a central $s_c$ with a given $T_i/T_F$, we calculated for different $T$ inside the lattice the entropy profiles and the total entropy $S_{tot}/L=\pi^2k_B T_i/T_F$.
}
\vspace*{-0.3cm}
\label{fig:3}\end{figure}

The spatially-dependent entropy distribution opens interesting perspectives for adiabatic spin cooling. 
A possible scheme would consist on three steps.
On a first stage a two-component balanced mixture is created at the lowest possible temperature $T_i$,
being stabilized against spin-changing collisions~\cite{Ho1998} by means of a sufficiently large homogeneous QZE. 
On a second stage a lattice is adiabatically grown and
the homogeneous magnetic field is adiabatically decreased allowing for
spin-changing collisions throughout the sample. The drop of the temperature in the lattice with the adiabatic decrease of the homogeneous magnetic field
can be estimated from the isentopic curves on Figure~\ref{fig:Isentropes}. However, to enter the spin coherent regime local entropy should be reduced. This is achieved in the final step which consists on slowly changing the
QZE into a non-uniform profile by
means of microwave or optical techniques~\cite{Tannoudji1972,Gerbier2006,Santos2007}, leading to the coexistence
of a spin-$1/2$ Heisenberg antiferromagnet (HAF) at the trap center and a spin-$S$ spin liquid at the
wings~(Fig.~\ref{fig:2}).
Figure~\ref{fig:3} shows, as a function of the initial $T_i/T_F$ ($T_F$ is the Fermi temperature of the 
original spin-$1/2$ prior to the lattice loading, see below), the central entropy per particle with a high homogeneous QZE (dashed line) 
and with a Gaussian QZE (solid line). For the case considered 
$s/k_B=0.35$ is achieved for the homogeneous case at $T_i/T_F\simeq 0.035$ whereas for the inhomogeneous QZE profile 
it is reached at $T_i/T_F\simeq0.09$, showing that an inhomogeneous QZE may allow for a 
large entropy reduction at the center as a result of the entropy excess at the storage wings. 
Note that for $T_i/T_F\simeq 0.1$ the central entropy may be reduced by more than factor of 2 
with respect to the entropy expected for the homogeneous spin-$1/2$ case.


Let us briefly comment on higher dimensional ($d=2,3$) systems. A simple estimate may be obtained from a spin-wave analysis when $a_2 > a_0$, 
where at least in the limit $a_2\gg a_0$ N\'eel order sets in on bipartite lattices~\cite{Troyer}, however due to frustration the modulus of the N\'eel order parameter decreases with increasing spin. There is just one spin-wave mode at $q\to \infty$ 
and three spin-wave modes at $q=0$. Taking the values of spin-wave velocities from~\cite{KolezhukVekua}, we obtain that at $T\to 0$ the factor $\gamma$ is given by Eq.~(\ref{gain}), albeit with $N(N-1)/2$ changed into $1+2[(a_2+a_0)/(a_2-a_0)]^d$, implying that \textit{the cooling should be even more efficient for 
higher dimensions}. Note that the spin-wave analysis predicts $\gamma$ to increase indefinitely when approaching $a_2=a_0$, although spin-wave 
analysis becomes less reliable in the vicinity of that point~\cite{KolezhukVekua}.

\section{Experimental feasibility}

A possible experimental sequence may be devised for e.g. $^{40}$K. Prior to the lattice
loading, a balanced mixture of e.g. $F=9/2, M_F=-5/2$ and $F=9/2, M_F=-7/2$ states is prepared 
at the lowest possible $T_i/T_F$ using standard techniques. At this stage a sufficiently large homogeneous magnetic
field guarantees that the initial two-component
mixture is stable against spin-changing collisions. The gas can be approximated with a
good accuracy by a free Fermi gas and therefore the initial entropy
per particle is given by $s=k_B\pi^2T_i/T_F$~\cite{Carr2004}. State of the art experiments may reach
at this stage $Ti/T_F \approx 0.1$, corresponding to an entropy per particle of $s \approx k_B$. Once
the gas is cooled down, a 3D lattice is grown and under proper conditions a Mott
insulator with one particle per site develops at the trap center~\cite{Jordens2008,Schneider2008}.
The next step consists in slowly lowering the magnetic field to allow for quasi-resonant spin-changing
collisions~\cite{Bornemann2008} throughout the
sample leading to a redistributed population (between $F=9/2$, $M_F=-3/2$, $-5/2$, $-7/2$, and $-9/2$~\cite{footnote-spchcolls}) and a significant drop of $T$
in the lattice. Temperature drop after this stage for 1D case, (assuming adiabaticity) is depicted on  Fig. 3. Note that fermions with $S>1/2$ suffer of large three-body loses~\cite{Ottenstein2008}. However, in our scheme $S>1/2$ is explored only at later stage, when the system is already in the hard-core regime with one fermion per site, thus three-body loses are largely suppressed.

Finally, the $F=9/2, M_F=-3/2$ and $F=9/2,M_F=-9/2$ states may be slowly expelled from the
trap center by the use of two Raman beam pairs. These beams couple $F=9/2$, $M_F = -3/2$
to $F=7/2$, $M_F=-1/2$ and $F=9/2$, $M_F = -9/2$ to $F=7/2$, $M_F=-7/2$. This
$\Delta M_F = -1$ coupling may be realized with just three lasers. A slight blue
detuning ensures that the resulting dressed states experience a repulsive potential. The
effect on the $M_F=-7/2$, $-5/2$ states is largely suppressed by the ratio
between Raman detuning and Zeeman splitting, which is kept on the order of $q_{cr}$.
Furthermore, the Raman beams shift the resonance condition for spin-changing collisions
in the center of the trap such that no more atoms in $M_F=-9/2$ and $M_F=-3/2$ are produced.
The outer regions have then a larger effective spin and hence act as
entropy absorbers, as discussed above.
Note that, interestingly, the center of the Raman beams acts as a crystallization point
for a slowly-growing N\'eel state (if $s_N$ is reached). 
In this way the problem of approaching the ground state despite the presence of many metastable low energy states may be
circumvented.

Let us mention some final remarks. For simplicity of our calculations we have considered 
cooling only within a Mott insulator. In the presence of particle-hole excitations we expect an even larger cooling efficiency~\cite{Schneider2008}. 
Note that the $S=1/2$ region is deeper in the Mott region 
than the $S>1/2$ region, and hence the $S>1/2$ region is expected to contain a larger particle/hole entropy. We stress also that the cooling efficiency is based on the assumption of adiabaticity. Spin-changing collisions are crucial in our cooling scheme. Although they are low energetic, they will be typically faster than super-exchange, and hence the adiabatic requirements of our model are comparable to those in other cooling schemes based on entropy relocations in optical lattices.

\section{Conclusions}

In summary, we studied a possible route for adiabatic spin cooling using high-spin lattice fermions.
The process resembles demagnetization cooling, but the role of the magnetic field is played by a spatially dependent 
QZE, and spin flip is substituted by spin-changing collisions. The spatially dependent QZE 
leads to two distinct regions of different 
effective spin ($S=1/2$ and $S>1/2$). At high $T$ the outer spin-$S$ region 
acts as an entropy absorber simply due to the larger spin, whereas the same remains true at even very 
low $T$ due to frustration. As a result we showed that a significant reduction of the entropy 
(more pronounced at lower $T$ and higher dimensions)
of the spin-$1/2$ Mott central region can take place.
Magnetic refrigeration using spatially variable QZE can hence significantly facilitate 
experimental realization of distinct
magnetic ground states, and in particular N\'eel ordering in spin-$1/2$ fermions.

{\it Acknowledgements.}

We thank A. Seel, J. Grelik and U. Schneider for discussions. We acknowledge support from the Center of Excellence QUEST, the 
ESF (EuroQUASAR), and the SCOPES Grant IZ73Z0-128058.


\begin{thebibliography}{10}
\bibitem{footnote1} In 1D and 2D, it must be understood as the characteristic temperature 
where e.g. the specific heat presents a peak.
\bibitem{footnote2} Even if N\'eel order eventually exists, due to the high frustration the ordering temperature will be much lower than the N\'eel temperature of the spin-$1/2$ antiferromagnet.
\end{thebibliography}
\end{document}